\documentclass{article}
\usepackage[utf8]{inputenc}
\usepackage[margin=1in]{geometry}
\usepackage{soul,xcolor}
\usepackage{caption,subcaption}
\usepackage{authblk}
\usepackage{lineno}

\usepackage[square,numbers,sort&compress]{natbib}
\usepackage{enumitem,amssymb,graphicx}
\newlist{todolist}{itemize}{2}
\setlist[todolist]{label=$\square$}
\usepackage{pifont,placeins}

\usepackage{hyperref}
\hypersetup{
    colorlinks=true,
    linkcolor=blue,
    filecolor=magenta,      
    urlcolor=blue,
    pdftitle={Overleaf Example},
    pdfpagemode=FullScreen,
    citecolor=blue,
    }

\usepackage[noabbrev,nameinlink]{cleveref}

\title{Graph Convolutional Neural Networks as Surrogate Models for Climate Simulation}
\author[1]{Kevin Potter}
\author[1]{Carianne Martinez}
\author[1]{Reina Pradhan}
\author[1]{Samantha Brozak}
\author[1]{Steven Sleder}
\author[1]{Lauren Wheeler}
\affil[1]{Sandia National Laboratories, Albuquerque, NM, USA}

\begin{document}

\maketitle

\begin{abstract}

Many climate processes are characterized using large systems of nonlinear differential equations; this, along with the immense amount of data required to parameterize complex interactions, means that Earth-System Model (ESM) simulations may take weeks to run on large clusters. Uncertainty quantification may require thousands of runs, making ESM simulations impractical for preliminary assessment. Alternatives may include simplifying the processes in the model, but recent efforts have focused on using machine learning to complement these models or even act as full surrogates. \textit{We leverage machine learning, specifically fully-connected neural networks (FCNNs) and graph convolutional neural networks (GCNNs), to enable rapid simulation and uncertainty quantification in order to inform more extensive ESM simulations.}
Our surrogate simulated 80 years in approximately 310 seconds on a single A100 GPU, compared to weeks for the ESM model while having mean temperature errors below $0.1^{\circ}C$ and maximum errors below $2^{\circ}C$.
    
\end{abstract}

\section{Introduction}


As global temperatures continue to rise, the need for effective and systematic evaluation of climate intervention strategies becomes increasingly important. Stratospheric Aerosol Injection (SAI) is one such strategy and like all brings significant risks \cite{sai-cairns2014security, sai-versen2021preparing} necessitating careful planning and evaluation of the positive and negative impacts.  
The Performance Assessment (PA) framework, a methodology originally designed for nuclear waste management \cite{PA-marietta2011sandia}, can be applied to the assessment of climate intervention strategies.  The Performance Assessment for Climate Intervention (PACI) framework\cite{PACI-10.3389/fenvs.2023.1205515} adapts the PA methodology to evaluate SAI by establishing a set of performance goals, identifying relevant system features, events, and processes (FEPs), and assessing the system’s performance, including uncertainties, against these goals.

The PACI framework aims to provide a structured and quantifiable approach to evaluate the risks and benefits of SAI in comparison to other climate pathways. Using the GLENS scenario, PACI demonstrates the application of PA to climate intervention by setting performance goals for a range of climate model output variables. The research highlights the importance of developing robust assessment tools to support informed decision-making in the context of climate intervention, while also acknowledging the need for future work to refine and expand the framework to address uncertainties and improve its applicability across different scenarios and regions.

However, a comprehensive PA, that demonstrates that all possible impacts have been investigated requires thousands or more simulation models of not only earth system models but regionally refined simulation runs to properly quantify the uncertainty in the predictions. Traditional ESMs can take several days to weeks to run on a large clusters \cite{acosta_2023,wang_2011} and quickly become too expensive to provide the necessary runs to address all possible impacts. To that end, simplified models \cite{nakatsugawa_1996,bravar_1991,khodayari_2013} or surrogate models \cite{kashinath_2021,dueben_2018} that can give ESM-like results in a shorter time scale (even if at a lower fidelity) are necessary for PA to be successfully applied. 

The application of neural networks (NNs) as surrogate models in climate science offers a promising approach to addressing the computational challenges associated with traditional climate models. These conventional models, while highly detailed and accurate, are computationally intensive and time-consuming to run. NNs, on the other hand, can be trained to emulate the behavior of these models, providing a faster and more efficient means of conducting climate simulations and analyses. By leveraging large datasets and the powerful learning capabilities of NNs, researchers can achieve faster predictions and insights, facilitating more responsive and informed decision-making in climate policy and research. This integration of NNs into climate science not only enhances the efficiency of modeling efforts but also opens new avenues for exploring the dynamics of our changing climate.

NNs have become a standard approach in the field of artificial intelligence for modeling otherwise difficult-to-capture nonlinear relationships. In the context of climate science, these networks serve as potent surrogate models, capable of approximating complex climate systems with remarkable accuracy. NNs consist of layers of interconnected neurons, where each neuron in one layer is connected to every neuron in the subsequent layer. This architecture allows NNs to capture and learn from the multifaceted interactions within climate data, making them particularly suitable for tasks such as climate prediction, trend analysis, and scenario simulation.

In this work, we use fully connected neural networks (FCNNs) and graph convolutional neural networks (GCNNs).  FCNNs, sometimes referred to as multi-layer perceptrons (MLPs), consist of multiple layers of neurons that take a fixed size vector as input and output a fixed size vector.

GCNNs presume a graph structure to the input where the data takes the form of nodes and edges (connections between nodes).  Different types of convolutional layers may require additional information, like position or specific edge attributes.  GCNN layers use a common set of weights to make predictions for each node utilizing the information from the node, its neighbors, and possibly edge attributes.  This is an increase in complexity from FCNNs as now additional information must be tracked in the form of edges and neighbors. However, the improved capability of understanding the local context and resulting accuracy improvements may justify the additional complexity cost.

\section{Model architecture}

\subsection{Fully-connected Neural Network (FCNN)}
\begin{figure}[]
    \centering
    \includegraphics[width=0.5\textwidth]{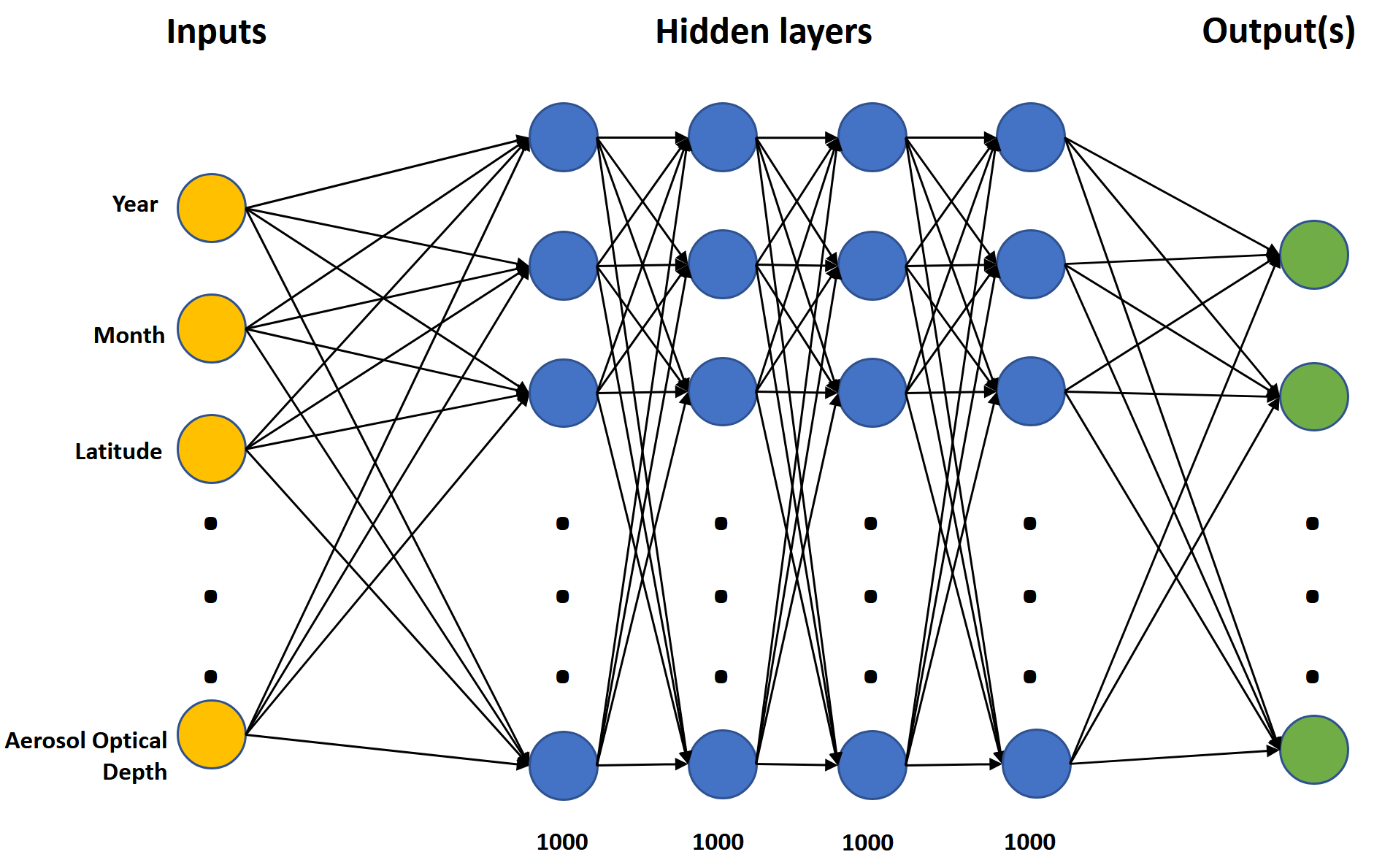}
    \caption{Diagram illustrating the architecture of an FCNN, showing the full connections between layers.}
    \label{fig:fcnn_model}
\end{figure}

A fully-connected neural network is one in which each neuron is connected to all neurons in the previous layer (illustrated in Figure \ref{fig:fcnn_model}), as opposed to a convolutional network where neuron connections are local and thus sparser.

\subsubsection{FCNN architecture details}

The FCNN in this work contains four linear layers with 1000 neurons each, uses a ReLU activation function followed by batch normalization layers. Finally, dropout is applied on each layer. Dropout was incorporated into the hidden layers of the network to improve regularization and quantify uncertainty. Dropout is a popular regularization technique to avoid overfitting where neurons in the network are turned off with some probability $0<p<1$. Keeping these layers active during inference approximates a Gaussian process \cite{gal_2016}. Hence, we can generate an ensemble of predictions and use these predictions to quantify uncertainty. The dropout rate in this work was set to 30\%, or $p=0.3$. During inference, 48 predictions were generated for each point and the uncertainty was characterized using standard deviation of those 48 predictions.

\subsection{Graph Convolutional Neural Network (GCNN)}

\begin{figure}[]
    \centering
    \begin{subfigure}[]{0.4\textwidth}
        
        \includegraphics[width=\textwidth]{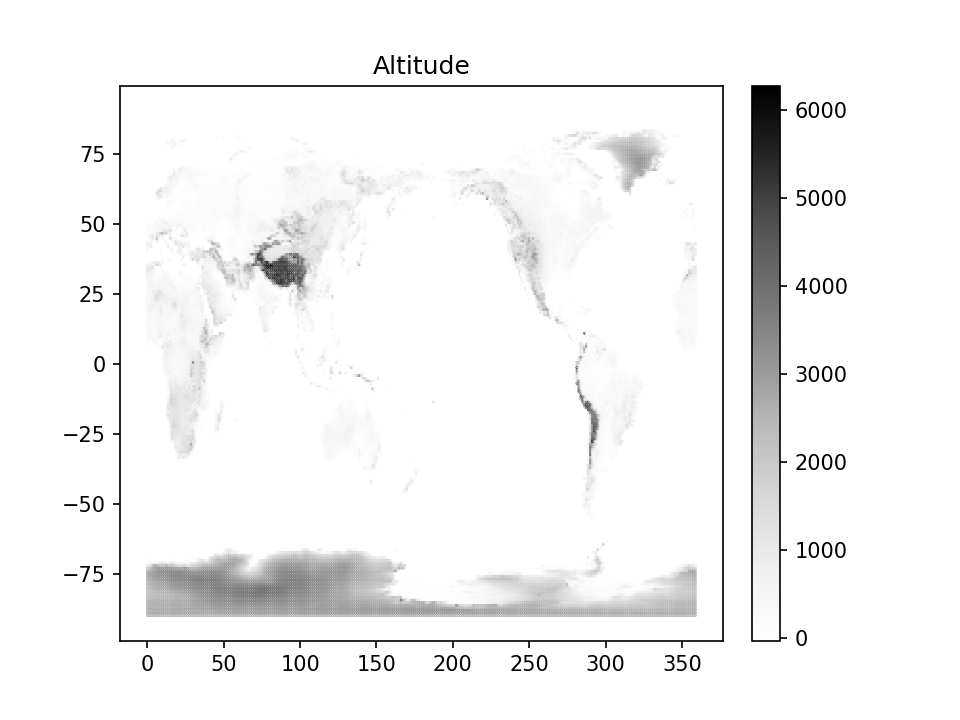}
        \centering
        \caption{}
        \label{fig:altitude}
    \end{subfigure}
    \begin{subfigure}[]{0.4\textwidth}
        
        \vspace{16pt}
        \includegraphics[width=\textwidth]{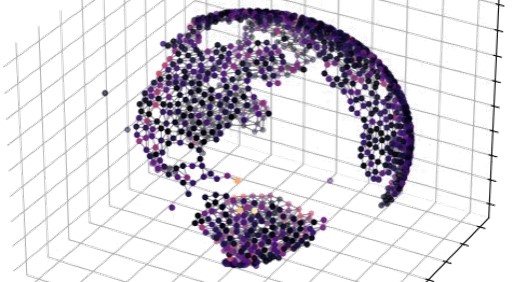}
        \vspace{8pt}
        \centering
        \caption{}
        \label{fig:graphs}
    \end{subfigure}
    \caption{Left: 2-D map projection of altitudes. Right: graph-structured downsampled map (land only).}
    \label{fig:maps}
\end{figure}
We chose a GCNN approach over a CNN approach for two reasons. First, two-dimensional projections of the Earth are prone to distortions, which could affect the results of more traditional convolutional neural networks.  Second, GCNNs allow the east to connect to the west naturally. \Cref{fig:altitude} shows over-sampling of the poles in this map projection. The graph structure naturally handles Earth's spherical geometry without the same distortions, as shown in the right panel of \Cref{fig:graphs}. Each node of the graph corresponds with a location on the Earth and contains information about the input variables as well as target variables for prediction. Edges connect the nodes, and graph neural networks propagate information along the edges of the graph. Edges were not part of the original data and we created them by connecting each node to its 4 nearest neighbors.

\begin{figure}
    \centering
    \includegraphics[width=0.9\textwidth]{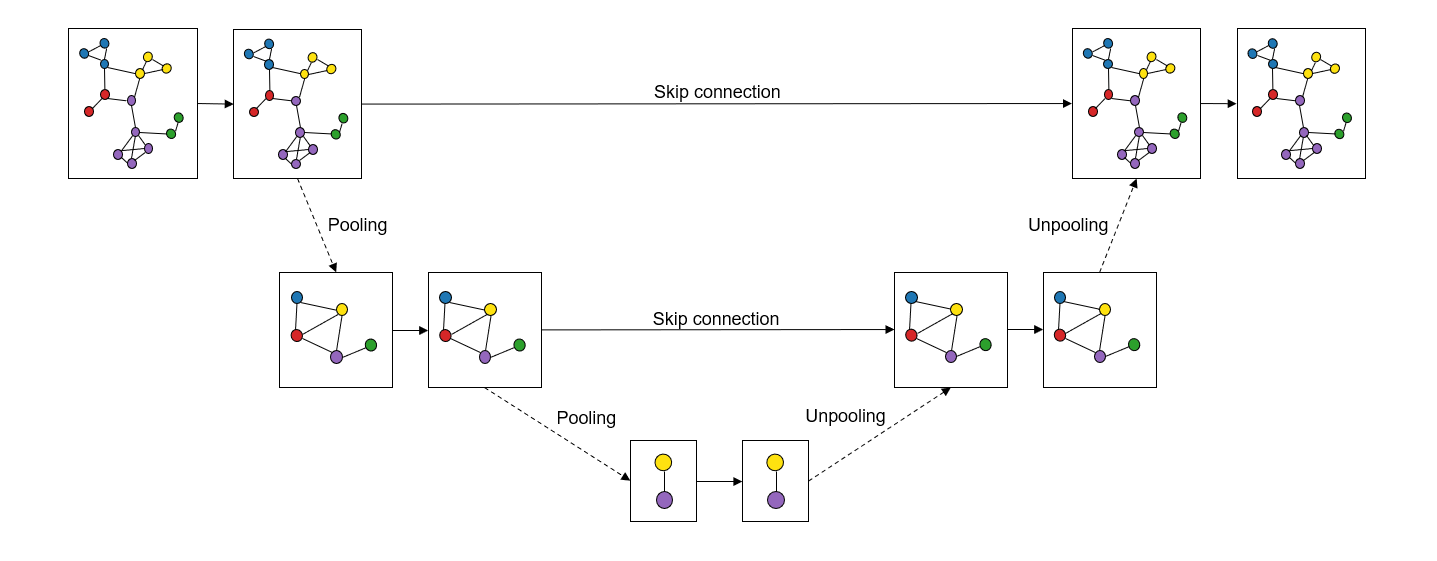}
    \caption{Diagram illustrating the architecture of a UNet for GCNNs.}
    \label{fig:gcn_unet}
\end{figure}

We chose a UNet-style \cite{unet10.1007/978-3-319-24574-4_28, gao2019graph-unet} for the GCNN architecture (basic example shown in \Cref{fig:gcn_unet}).  UNet uses successive coarsening operations intermixed with convolutions.  This downscaled graph is then upsampled to the same graph layout as prior to downsampling.  This allows skip connections between the state of the network prior to coarsening to be passed to the upsampled operation and concatenated.  This is hypothesized to allow the network to retain fine details while simultaneously getting information from a larger context.

\subsubsection{GCNN architecture details}
The network had 4 downsampling operations using voxel grid pooling for coarsening.  At each downsampling, the voxel grid size is doubled (such that we end with a $30\times30\times30$ grid at the deepest layer) and the number of features per node is doubled (starting from 64 features).  Each downsample performs the convolutions, a ReLU activation, and then voxel grid pooling.  Each upsample performs the convolution, 2 linear (NN layer operating on each node separately) operations with ELU activations \cite{elu-clevert2015fast}, and then a KNN interpolation to the upsampled graph with a k of 10.  Finally there are 2 convolutions with SeLU activations \cite{selu-klambauer2017self} and a final linear operation to get to the output.

\subsection{Data and training}
The original GLENS data was converted into csv (comma separated value) files. This allowed easier use in both the FCNN and the GCNN models. 
\vspace{10pt}

\noindent The FCNN model was trained to predict the difference in TSA between control and feedback runs. The input variables were: year, month, latitude, longitude, and AODVIS (aerosol optical depth, 550 nm). 
\vspace{10pt}

\noindent The GCNN model was trained on four control runs under the RCP8.5 scenario obtained from the Geoengineering Large Ensemble Project (GLENS) \cite{tilmes_2018}. The input variables were: month, latitude, altitude, and time $t$. Time is given by the year plus twelve times the month. Longitude was omitted in order to reduce overfitting. Latitude and longitude were converted to an x, y, z position triplet (for a radius of 1).

The predicted (output) variables were given by:

\begin{itemize}
    \item TSA: Two-meter air temperature (land only).
    \item TREFHT: Reference height temperature.
    \item PRECT: Total (convective and large-scale) precipitation rate (liquid + ice).
    \item ICEFRAC: Fraction of surface area covered by sea ice.
    \item SNOWHLND: Water-equivalent snow-depth.
    \item ALTMAX: Maximum annual active layer thickness.
\end{itemize}

Because TSA does not have values over the ocean, missing TSA values were replaced by the TREFHT at the equivalent location. Missing values for ALTMAX were replaced with 35.1776 m (ALTMAX has a bimodal distribution and so this value was chosen because that was at least 75\% of the values). All other missing values were replaced by zeros. The PRECT variable was rescaled from the average rainfall in cm per second to average rainfall in cm per day. The previous month's output data was also included to allow for autoregressive prediction. Both models were trained for 100 epochs with a batch size of one timestep (all nodes for a particular month).

\section{Results}
\subsection{FCNN}
\begin{figure}[]
    \centering
    \includegraphics[width=\textwidth]{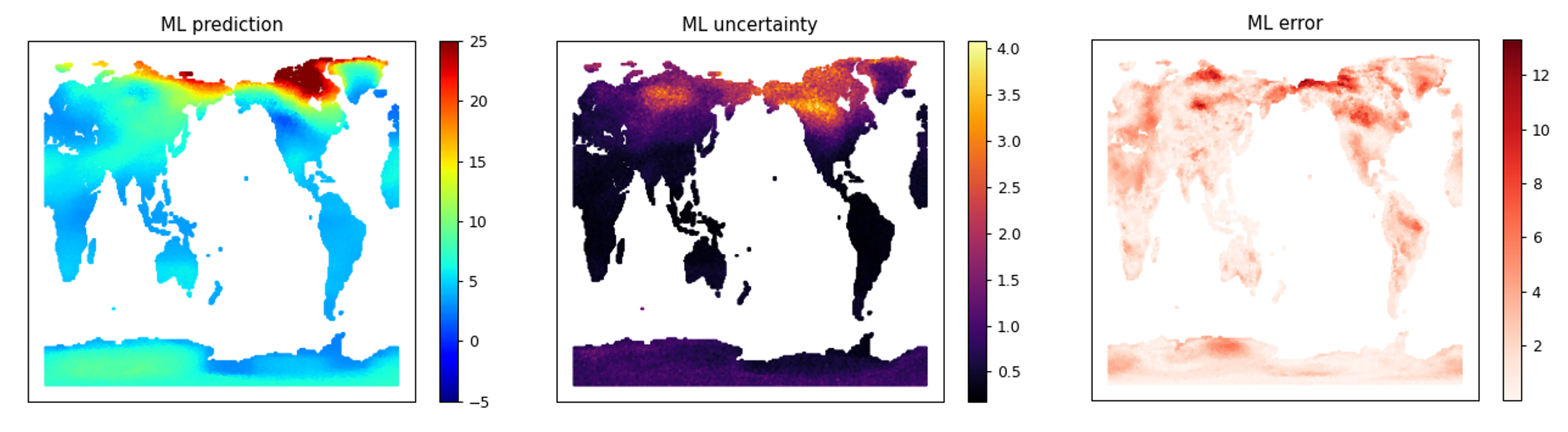}
    \caption{Left: FCNN prediction for the difference in TSA between a control and feedback run for one month (January 2080). Center: uncalibrated uncertainty estimates for FCNN prediction. Right: FCNN error with respect to GLENS simulation data.}
    \label{fig:fcnn_results}
\end{figure}

Dropout improved mean absolute error on the test set when compared to the no dropout regime (\Cref{tab:fcnn_table}). When examining the map projection in \Cref{fig:fcnn_results}, the uncertainty qualitatively aligns with the error in the northern hemisphere. However, the uncertainty does not appear to be correlated with the FCNN's error in South America, Africa, and parts of Antarctica. This work did not attempt to calibrate uncertainty estimates as formal uncertainty quantification was out of scope for this project. Dropout provides a quick mechanism to characterize FCNN prediction uncertainty, but future work is needed to understand its reliability for practical use in this domain.

\begin{table}[]
    \centering
    \caption{Comparison of mean absolute error (MAE) between FCNNs with and without dropout.}
    \label{tab:fcnn_table}
    \begin{tabular}{lll}
        & Without dropout & With dropout \\
        \hline
        Train     & \textbf{1.070}            & 1.638                     \\
        Test      & 2.608                     & \textbf{2.182}                    
    \end{tabular}
\end{table}

\subsection{GCNN}
We developed BlockNet, a Python package which streamlines the implementation, training, and testing of a graph convolutional neural network. BlockNet includes data processing and methods for logging the training process.  We implemented the graph model using PyTorch-Geometric \cite{fey_2019}.

The GCNN could simulate 80 years in approximately 310 seconds on a single A100 GPU, which is remarkably faster than the typical ESM model which take weeks on a high capacity cluster.

\begin{figure}[]
    \centering
    \begin{subfigure}[b]{0.9\textwidth}
             \caption{}\vspace{-2pt}
             \includegraphics[width=\textwidth]{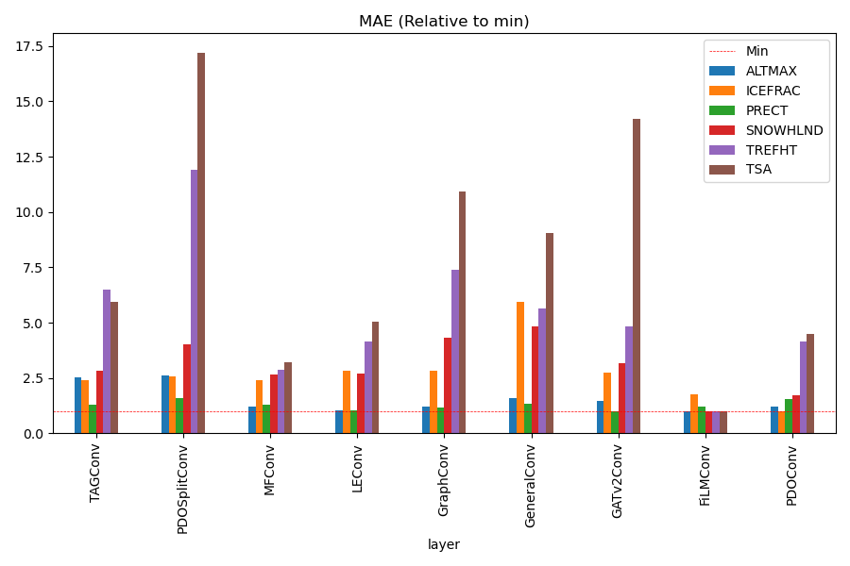}
             \label{fig:layer_MAE}
    \end{subfigure}
    \begin{subfigure}[b]{0.9\textwidth}
             \caption{}\vspace{-2pt}
             \includegraphics[width=\textwidth]{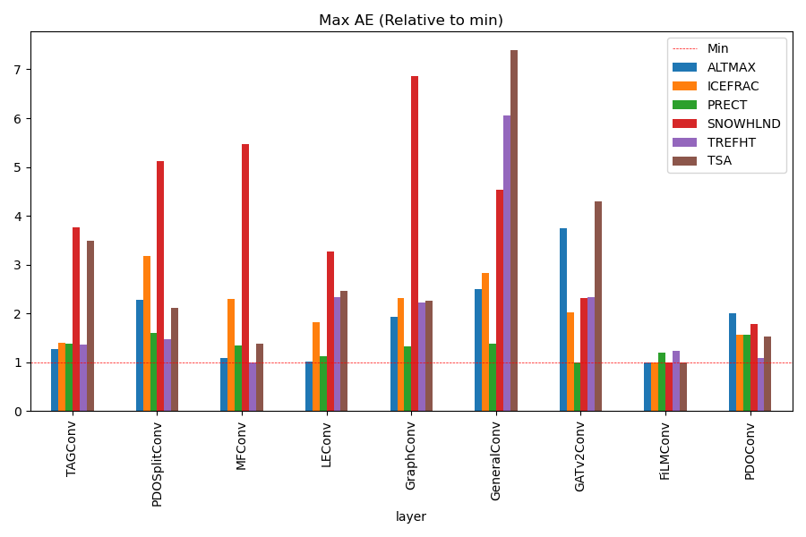}
             \label{fig:layer_RMAE}
    \end{subfigure}
    \caption{Layer tests for the GCNN in terms of mean absolute error (MAE) and maximum absolute error (MaxAE). Charts are normalized by the minimum MAE and MaxAE for comparison.}
    \label{fig:layertests}
\end{figure}

Extensive testing was done in order to determine the best performing type of layer for the GCNN. Twenty types of layers were tested using the same training procedure described previously, and results for the best nine are shown in \Cref{fig:layertests}. The FiLMConv (convolutional feature-wise linear modulation) layer  \cite{brockschmidt_2020} performed the best across most of predicted variables in terms of both MAE and Max AE and remained competitive for those where it was bested.

\begin{figure}[]
    \centering
    \begin{subfigure}[b]{0.4\textwidth}
             \caption{}\vspace{-2pt}
             \includegraphics[width=\textwidth]{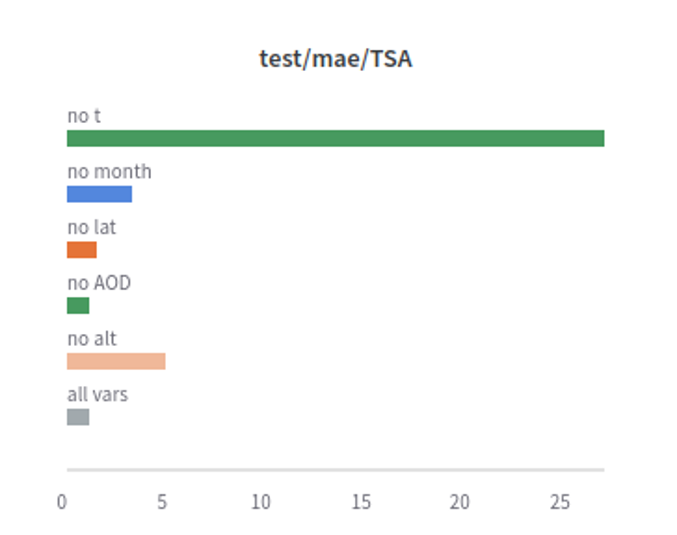}
             \label{fig:TSA_varimp}
    \end{subfigure}
    \begin{subfigure}[b]{0.4\textwidth}
             \caption{}\vspace{-2pt}
             \includegraphics[width=\textwidth]{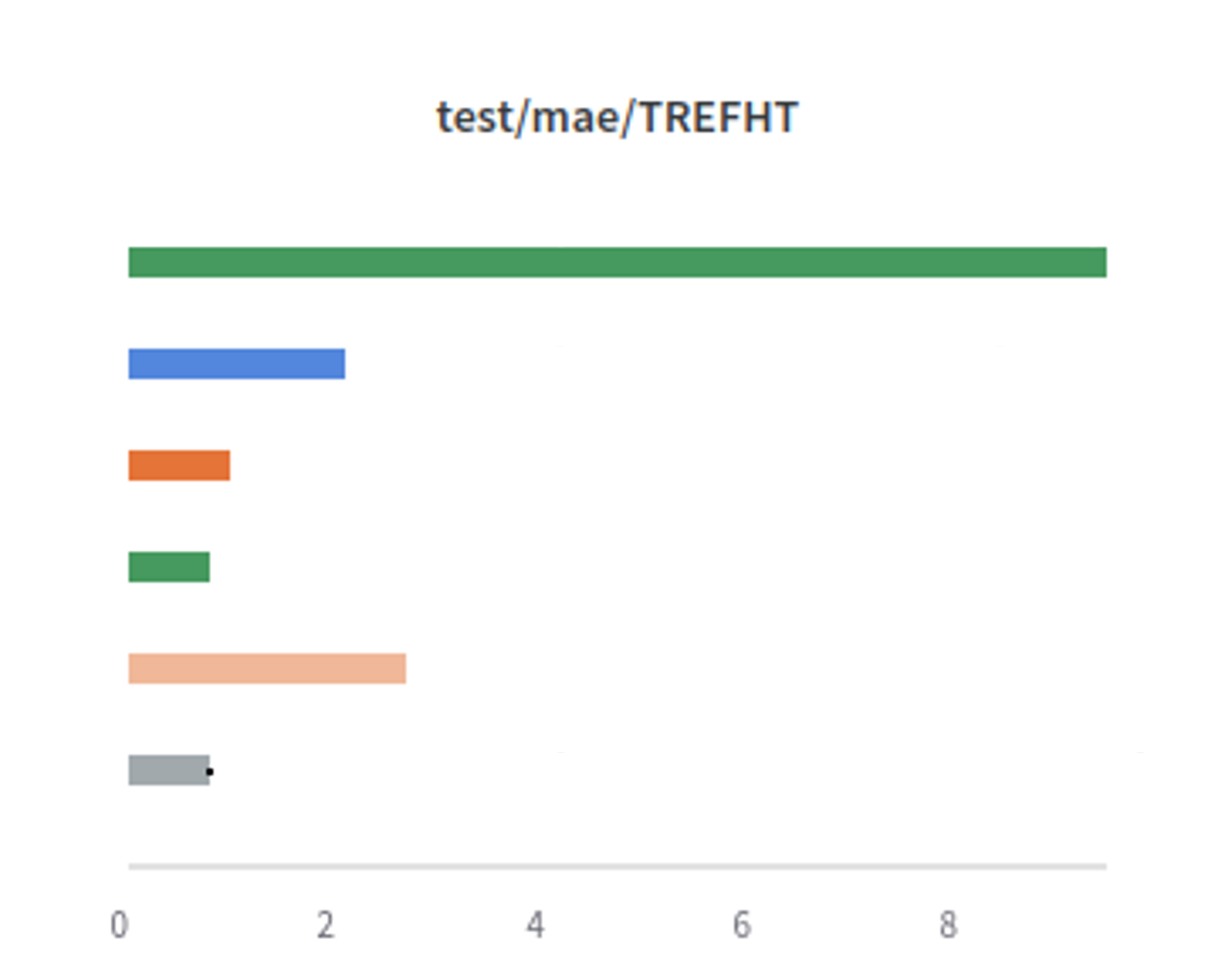}
             \label{fig:TREFHT_varimp}
    \end{subfigure}
    \caption{Variable importance tests for the GCNN. Other variables reported similar results and are omitted. The horizontal axis is mean absolute error (MAE).}
    \label{fig:varimp}
\end{figure}

Variable importance tests shown in \Cref{fig:varimp} reveal that the variable $t$ affects results the most, which is unsurprising given that $t$ is the only variable that has a relationship with CO$_2$ (At the time of testing, CO$_2$ had not been converted from GLENS). Altitude was the next most important, followed by month which was unexpected given that the month is recoverable from $t$.

\begin{figure}[]
     \centering
     \begin{subfigure}[b]{0.24\textwidth}
         \centering
         \caption{}\vspace{-2pt}
         \includegraphics[width=\textwidth]{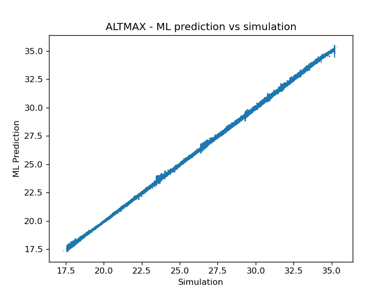}
         \label{fig:ALTMAX}
     \end{subfigure}
     \begin{subfigure}[b]{0.24\textwidth}
         \centering
         \caption{}\vspace{-2pt}
         \includegraphics[width=\textwidth]{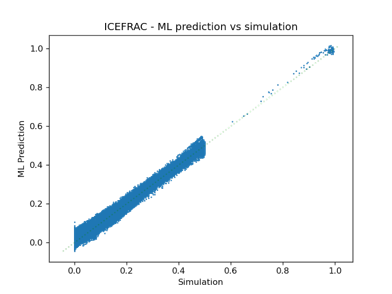}
         \label{fig:ICEFRAC}
     \end{subfigure}
     \begin{subfigure}[b]{0.24\textwidth}
         \centering
         \caption{}\vspace{-2pt}
         \includegraphics[width=\textwidth]{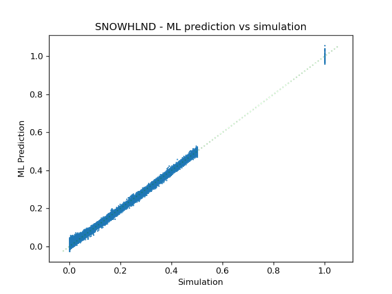}
         \label{fig:SNOWHLND}
     \end{subfigure}
          \begin{subfigure}[b]{0.24\textwidth}
         \centering
         \caption{}\vspace{-2pt}
         \includegraphics[width=\textwidth]{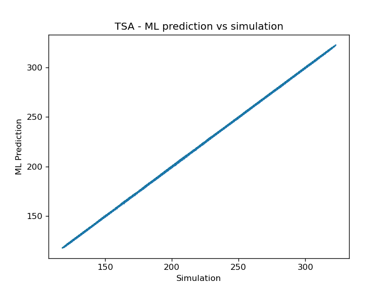}
         \label{fig:TSA}
     \end{subfigure}
        \caption{Plots showing the simulated GLENS output against the GCNN prediction. TSA comparisons are done over land.}
        \label{fig:predvsim}
\end{figure}

\begin{figure}[!htbp]
     \centering
     \begin{subfigure}[b]{0.38\textwidth}
         \centering
         \caption{}\vspace{-2pt}
         \includegraphics[width=\textwidth]{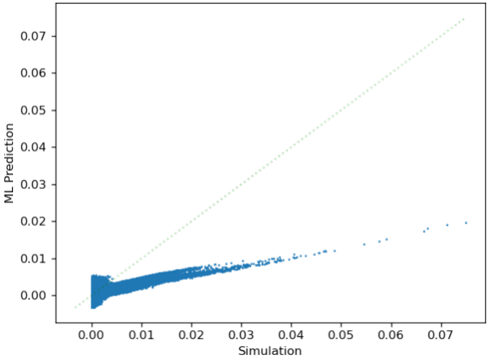}
         \label{fig:PREC_predvtrue}
     \end{subfigure}
     \begin{subfigure}[b]{0.4\textwidth}
         \centering
         \caption{}\vspace{-2pt}
         \includegraphics[width=\textwidth]{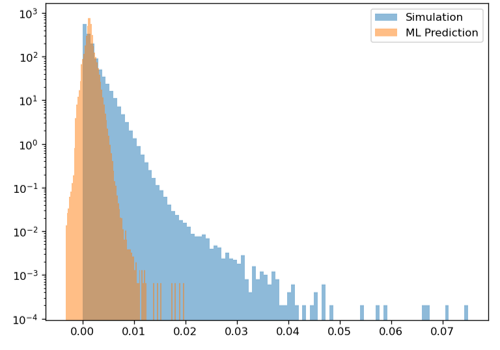}
         \label{fig:PREC_dist}
     \end{subfigure}
     \caption{Plots of true vs. predicted precipitation and distributions of true and predicted precipitation.}
\end{figure}

In terms of MAE, MARE, MaxAE, and MaxARE, the GCNN performed consistently better than the FCNN as shown in \Cref{tab:results_table}, with the exception of PRECT in which performance was comparable. The GCNN performance is also visualized in \Cref{fig:predvsim} for ALTMAX, ICEFRAC, SNOWHLND, and TSA. The linear trends indicate that the model is able to predict the data accurately. TSA has the least variance (\Cref{fig:TSA}) while ICEFRAC has the most variance (\Cref{fig:ICEFRAC}). However, \Cref{fig:PREC_predvtrue} clarifies how the GCNN struggled to predict PRECT. The model is consistently underestimating the true precipitation, and the residuals appear to be heteroskedastic with higher variance at lower levels of precipitation. A histogram of the precipitation predictions also shows that the model does not capture the distribution of the ground truth (\Cref{fig:PREC_dist}).

\begin{table}[]
    \centering
    \caption{Mean absolute error (MAE), mean absolute relative error (MARE), maximum absolute error (MaxAE), and maximum absolute relative error (MaxARE) for the fully-connected (FCNN) and the graph convolutional neural network (GCNN). TSA comparisons are done over land. Lower values are better for all metrics.}
    \label{tab:results_table}
    \begin{tabular}{llllllll}
        & \textbf{Model} & \textbf{ALTMAX} & \textbf{ICEFRAC} & \textbf{PRECT} & \textbf{SNOWHLND} & \textbf{TREFHT} & \textbf{TSA} \\
        \hline
            \textbf{MAE} & FCN            & 0.4611          & 0.1504           & 0.0012         & 0.1238            & 5.7008          & 21.7042      \\
            & GCNN      & {0.0341}         & {0.0019}          & {0.0010}           & {0.0038}         & {0.0707}            & {0.0837} \\
        \hline
            \textbf{MARE} & FCN            & 0.0163          & 131.2807         & 0.6694         & 107.9613          & 0.0409          & 0.1306       \\
            & GCNN      & {0.0010}         & {0.9655}          & {0.4658}           & {2.7821}         & {0.0005}            & {0.0005} \\
        \hline
            \textbf{MaxAE} & FCN            & 15.3309         & 1.3128           & 0.0736         & 0.8411            & 29.4854         & 84.0491      \\
            & GCNN      & {0.6211}         & {0.1658}          & {0.0697}           & {0.1883}         & {1.7796}            & {1.9909}  \\
        \hline
            \textbf{MaxARE} & FCN            & 0.8522          & 485.1506         & 3.3429         & 427.6383          & 0.2321          & 0.6003       \\
            & GCNN      & {0.0347}         & 86.1159         & 3.2700           & 126.1915       & {0.0171}            & {0.0126} \\
    \end{tabular}
\end{table}

\section{Discussion and future work}
We have shown that GCNNs can act as efficient surrogate models for ESMs, enabling more rapid simulations before turning to the higher fidelity models. The GCNN architecture is able to predict thousands of outcomes in the time of a single ESM run. Furthermore, testing was carried out to tune the GCNN and identify the best type of layer for the model as well as important input variables. The GCNN overall outperformed the FCNN and combination networks, but struggled with predicting precipitation, likely due to the scale of the data.

The results presented here are for models trained only on GLENS control runs, and future work could incorporate GLENS feedback runs. Additionally, incorporating additional input variables (such as CO$_2$) into the model could improve results or allow the model to react to actual (or hypothetical) CO$_2$ emissions scenarios without requiring additional ESM runs for training data.

\section*{Acknowledgements}
This paper describes objective technical results and analysis. Any subjective views or opinions that might be expressed in the paper do not necessarily represent the views of the U.S. Department of Energy or the United States Government. Sandia National Laboratories is a multimission laboratory managed and operated by National Technology \& Engineering Solutions of Sandia, LLC, a wholly owned subsidiary of Honeywell International Inc., for the U.S. Department of Energys National Nuclear Security Administration under contract DE-NA0003525.

\bibliography{ref}

\end{document}